\documentclass[aps,prl,superscriptaddress,twocolumn]
{revtex4-2}

\usepackage[utf8]{inputenc}

\usepackage{amsthm,amsmath,amssymb,amsfonts,bbm,mathtools}
\usepackage{caption}
\usepackage{xcolor}
\usepackage{graphicx}
\usepackage{appendix}
\usepackage[normalem]{ulem}
\usepackage{mathrsfs}
\usepackage{enumitem}

\usepackage[autostyle]{csquotes}

\usepackage[colorlinks=true,linkcolor=blue,urlcolor=blue,citecolor=blue,pdfusetitle]{hyperref}

\usepackage{mathtools}

\begin{document}

\title{Leveraging quantum statistics to enhance heat engines}
\author{Keerthy Menon}
\email{keerthy.menon@oist.jp}
\author{Thomas Busch}
\author{Thomás Fogarty}
\email{thomas.fogarty@oist.jp}
\affiliation{%
 Quantum Systems Unit,  Okinawa Institute of Science and Technology Graduate University, Onna, Okinawa 904-0495, Japan.
}%

\date{\today}

\begin{abstract}
A key focus of designing quantum thermal devices is the potential advantage that can be gleaned from genuine quantum effects when compared to classical devices. The recent experimental realization of the Pauli engine \cite{Koch2023}—where energy is extracted via changes in particle statistics as an alternative to conventional heat sources—has opened new avenues of research where quantum statistics can be considered as a thermodynamic resource. In this work we propose hybrid quantum heat engines which utilize additional strokes that change the single particle statistics between bosonic and fermionic descriptions during the cycle. To accomplish this we consider the 1D Lieb--Liniger gas whereby the s-wave interactions can be tuned between the non-interacting and the hard-core limit, which are described by bosonic and fermionic statistics respectively. We show that by suitably choosing where to implement these statistical strokes during an Otto-like cycle, the efficiency and work output can be significantly enhanced when compared to fully bosonic or fully fermionic engines. Furthermore, in the degenerate regime our engine can operate at the Carnot efficiency, due to the interplay between the different contributions of heat and work induced by the statistical strokes. Finally, we highlight how our thermodynamic cycles can realize other thermal operations, such as refrigerators, promising similar statistical enhancements for a wide range of temperatures. 

\end{abstract}

\maketitle
\section{Introduction}
The study of quantum heat engines is an important facet of expanding our understanding into how genuine quantum effects may influence thermodynamic behavior \cite{Myers2022rev,CANGEMI20241}. Of particular interest is the emergence of collective many-body effects that can boost engine performance~\cite{Jaramillo_2016,Niedenzu_2018} by leveraging quantum coherence~\cite{Camati_2019,Klatzow_2019}, tunable interactions~\cite{Adolfo_2019,Wang_2020,Myers_2021,Mohammed_2023,Watson2023, Nautiyal_2024} or quantum phase transitions~\cite{Campisi2016,Revathy2020,Fogarty_2021}. Cold atomic gases offer an ideal testbed to explore these topics experimentally due to a high degree of control over trapping potentials and interactions~\cite{Bouton_2021,Nettersheim_2022,Simmons23}. 
In these systems particle statistics can play an important role, with bosonic and fermionic working media offering different advantages depending on the exact form of the energy spectrum~\cite{Zheng2015,Chen_2018,Sur2023}. In bosonic systems there can be a collective enhancement due to permutation symmetry \cite{wat20,mye20}, while an engine operated in the Bose-Einstein condensate regime can achieve a performance boost due to a vanishing work cost in compressing the gas~\cite{mye22a,Simmons23,Eglinton_2023}.

This highlights how different quantum statistics can enhance engine performance depending on the single-particle spectrum and the systems temperature. However, to fully leverage these advantages, the engine cycle must involve switching between bosonic and fermionic statistics during its strokes. While altering particle statistics may seem challenging, it is feasible in specific systems. For instance, in two-component Fermi gases, the scattering length can be tuned along the BEC-BCS crossover, enabling a transition between two distinct statistical regimes. In the BEC limit, fermionic atoms form molecular bosons that undergo Bose-Einstein condensation and obey Bose-Einstein (BE) statistics. In contrast, at the BCS end of the crossover, long-range Cooper pairs emerge, which follow Fermi-Dirac (FD) statistics. An engine operating between these two regimes was suggested and demonstrated in~\cite{Koch2023}, where the gas was compressed in the BEC regime and expanded in the fermionic regime, resulting in enhanced work output due to the large Fermi pressure. This engine, known as the \textit{Pauli engine}, operated without traditional heat baths, thereby making it fully isentropic. Instead, it converted energy from the change in particle statistics into potential energy within a magneto-optical trap.

 In this work we study a related model where the  particle statistics can be changed via modifying the s-wave interactions, namely the 1D Lieb-Liniger model. Here, the non-interacting limit is described by an ideal Bose gas (BG), while for infinite repulsive interactions the bosons effectively fermionize in what is known as the Tonks--Girardeau (TG) limit \cite{TG1960}. While similar in concept, discussing an engine with this kind of working medium has a particular advantage over the BEC-BCS regime used for the Pauli engine, as molecular bosons do not need to be formed at the cost of a large binding energy, which severely impacts efficiency. 
Furthermore, in this work we consider the operation of a typical quantum heat engine cycle, whereby heat from external baths is converted into work, with the addition of quantum statistics strokes which can boost performance. 

All engine setups we discuss below implement a modified Otto cycle: movable boundary conditions compress and expand the gas during the two adiabatic strokes, and the system equilibrates with different heat baths during the isochoric heating and cooling strokes. The statistical strokes can therefore be added in two different ways. Firstly, they can be implemented simultaneously with the adiabatic compression and expansion strokes thereby changing the amount of work done during the cycle. Alternatively, they can be performed during the equilibration strokes, which would modify the amount of heat absorbed from or dissipated to the baths. In both cases we show that the total work output and the efficiency of the cycles can be significantly increased  when the system is in the degenerate regime, while any gain is lost at higher temperatures.

The manuscript is ordered as follows:  we first introduce the Lieb--Liniger model which forms the working medium of our engines. Next we discuss the two engine setups which incorporate changes in particle statistics and quantify their performance compared to traditional quantum Otto cycles. In the last section we discuss the operational regimes of other quantum thermal machines, such as refrigerators, heaters and accelerations. Finally we conclude and give an outlook for future work.

\section{Model}
We consider a quantum heat engine with a working medium consisting of a one-dimensional gas of $N$  bosons of mass  $m$, confined within a box potential of length  $L$. The particles interact via short-range contact interactions characterized by an effective coupling constant  $g$, which is expressed in terms of the rescaled interaction strength  $c = m g/\hbar^2$. Using natural units where  $\hbar = 2m$,  $c$  acquires the dimensions of inverse length. This system is described by the Lieb-Liniger model with hard-wall boundary conditions \cite{Batchelor_2005}, and its Hamiltonian is given by
\begin{equation}
    H_{LL}=-\sum_{i=1}^N \frac{\partial^2}{\partial x_i^2} + 2c\sum_{i<j}^N \delta(x_i-x_j).
    \label{eq:LLHamiltonian}
\end{equation} 
The first term represents the single-particle kinetic energy, while the second term accounts for the two-body interaction energy. 

While the Lieb--Liniger model is exactly solvable for arbitrary interaction strength  $c$ using the Bethe Ansatz~\cite{Bethe_1931,Lieb_1963a,Oelkers_2006,Jiang_2015,Franchini_2017}, here we focus on two specific interaction limits. In the non-interacting limit ($c = 0$) the system reduces to an ideal Bose gas (BG) and is described by the Bose-Einstein (BE) distribution $b_n = 1 /\left[\exp\left((E_n-\mu)/(k_BT)\right)-1\right]$ with entropy $S^B = -k_B \sum_{n=0}^\infty b_n \ln b_n - \left(1+ b_n\right) \ln \left( 1+b_n\right)$. Here $T$ is the temperature of the ensemble, $k_B$ the Boltzmann constant, $\mu$ the chemical potential and $E_n = n^2\pi^2/L^2$ the single particle energies of the box potential. We also consider the strongly interacting limit ($c = \infty$), which is known as the Tonks--Girardeau (TG) limit, where the strongly repulsive bosons mimic the behavior of ideal spin-polarized fermions. This \textit{fermionization} of the strongly interacting bosons has an important consequence, namely that the thermodynamics of the TG gas is identical to that of fermions \cite{lenard1966one,atas2017exact} which allows them to be described by the Fermi-Dirac distribution $f_n = 1 /\left[\exp\left((E_n-\mu)/(k_BT)\right)+1\right]$ with entropy $S^F = -k_B \sum_{n=0}^\infty f_n \ln f_n + \left(1- f_n\right) \ln \left( 1-f_n\right)$. The system can be driven between these two regimes by using Feshbach resonances~\cite{Inouye_1998, Olshanii_1998,Bergeman_2003} that can tune the interaction strength.

\section{Statistically enhanced thermodynamic engines}

In this section, we introduce two distinct engines running on Otto-inspired cycle~\cite{Rezek_2006,Henrich_2007,Allahverdyan_2008,Gemmer_2009} which take advantage of changes in particle statistics. The first cycle introduces the change in particle statistics during the adiabatic work strokes, realized through independent and simultaneous changes in particle interactions \cite{Adolfo_2019} and box length \cite{Chen_2018}. This has been investigated for small systems and can be expected to lead to an enhancement in the engine performance~\cite{Mohammed_2023}. The second cycle incorporates the change in particle statistics into the isochoric thermal strokes. We will show that the work output benefits from heating that arises both from the heat exchange with the bath and the change in particle statistics. 

\subsection{Statistically enhanced adiabatic strokes: A-engine}
\begin{figure*}[tb]
    \includegraphics[width=0.85\linewidth]{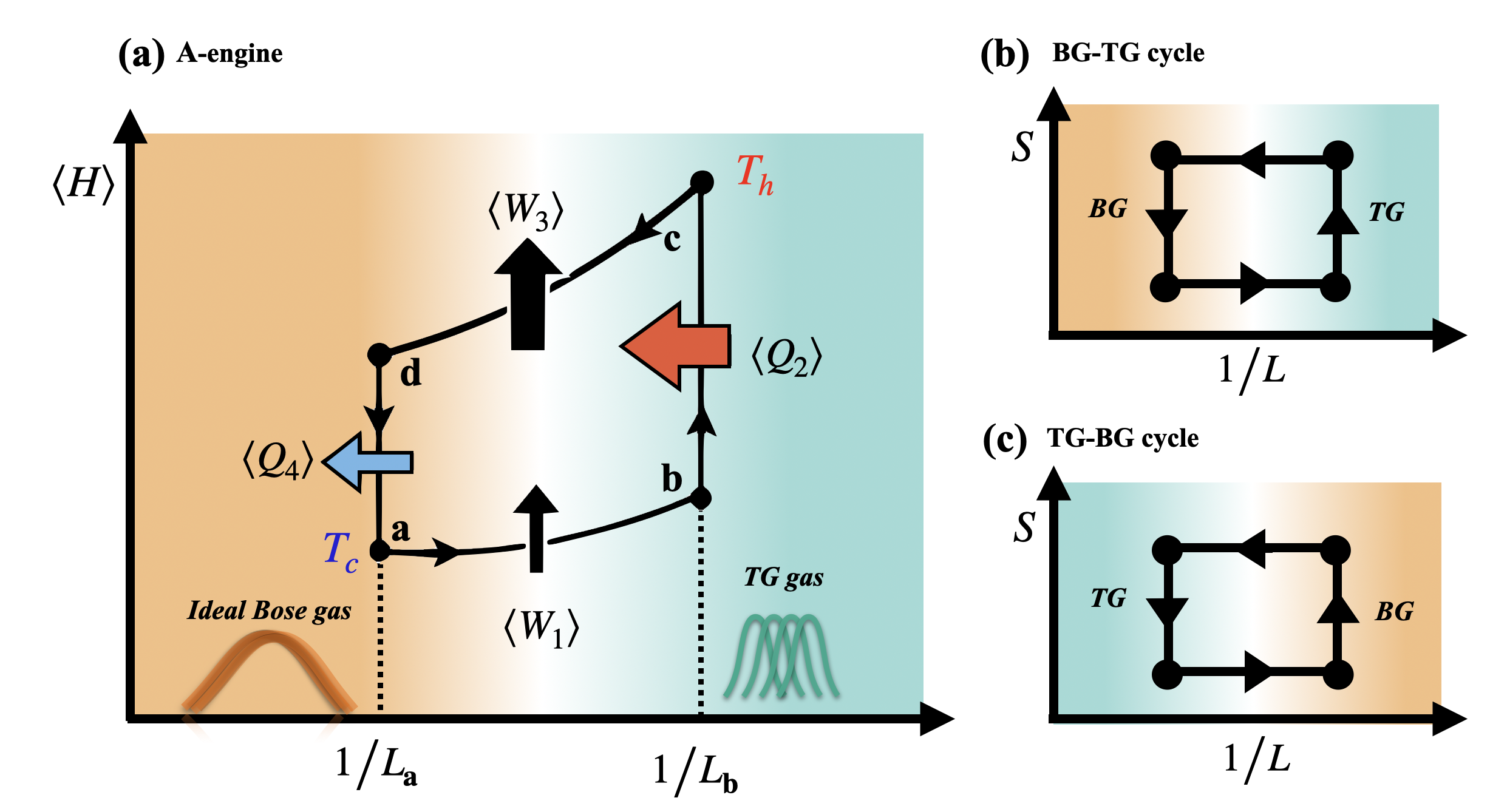}
    \caption{(a) Schematic of the \textbf{A}-engine, where the adiabatic strokes are statistically enhanced. During the work strokes, \textbf{a $\rightarrow$ b} and \textbf{c $\rightarrow$ d}, the length of the box and the particle statistics are adiabatically changed, and the work done is given by $\langle W_1 \rangle$ and $\langle W_3 \rangle$. In the TG regime (green region) the system is connected to a hot bath at temperature $T_h$, \textbf{b $\rightarrow$ c}, and heat the $\langle Q_2\rangle$ is absorbed, while in the ideal Bose gas regime (orange region) the system is coupled to a cold bath at temperature $T_c$ and heat $\langle Q_4\rangle$ is lost. The panels (b) and (c) show the change of the entropy, $S$, as a function of the inverse box length for the BG-TG cycle (first adiabatic stroke going from bosonic to fermionic statistics) and the TG-BG cycle (opposite case).}
    \label{fig:InteractionEngine}
\end{figure*}

Let us first discuss a modified Otto cycle where the change in particle statistics happens during the adiabatic strokes. We refer to this as the A-engine and the individual strokes of this cycle are shown in an energy-inverse volume diagram in Fig.~\ref{fig:InteractionEngine}(a), where the colored regions highlight the different particle statistics - orange for bosonic statistics ($c=0$) and green for fermionic statistics ($c\rightarrow\infty$). The individual strokes of the \textbf{A}-engine are as follows:

\noindent
\begin{enumerate}
    \item \textbf{a $\rightarrow$ b} \textit{Adiabatic Compression}:  the system is initially in thermal equilibrium with a cold reservoir at temperature $T_c$ at point \textbf{a}. The bosons are confined in a box potential of length $L_{\textbf{a}}$ and do not interact, $c = 0$. They are therefore described by the BE distribution. The system is then decoupled from the reservoir, and the box length is compressed from $L_{\textbf{a}}$ to $L_{\textbf{b}}$. Simultaneously, the interaction strength is increased to $c\rightarrow\infty$ and the system is described by the FD distribution at the end of the stroke at point $\textbf{b}$. As this is an adiabatic process the entropy must remain constant, $\Delta S = S^F-S^B = 0$, and therefore the TG gas must be described by a different effective temperature $T_c'$. Maintaining adiabaticity ensures that the system evolves without the exchange of heat and therefore only work is done on the system during this stroke. The amount of work done is given by $ \langle W_1 \rangle = \sum_{n}\left( f^{(\textbf{b})}_n(T_c') E^{(\textbf{b})}_n - b_n^{(\textbf{a})}(T_c)E^{(\textbf{a})}_n \right)$ with $E^{(\textbf{a})}_n \equiv E_n(L_{\textbf{a}})$ and $E^{(\textbf{b})}_n \equiv E_n(L_{\textbf{b}})$ being the single particle energies at the initial box length $L_{\textbf{a}}$ and final box length $L_\textbf{b}$, respectively. Here $b^{(\textbf{a})}_n(T_c)$ is the BE distribution at point \textbf{a} with temperature $T_c$, while $f^{(\textbf{b})}_n(T_c')$ is the FD distribution at point \textbf{b} with effective temperature $T_c'$. This stroke combines changes in the box length and interaction strength, resulting in a statistics-driven energy shift crucial for the engine’s operation.
    
    \item \textbf{b $\rightarrow$ c} \textit{Isochoric heating}: During this stroke, the system is brought into thermal contact with a hot reservoir at temperature $T_h>T_c$ and heat is exchanged between the system and the hot bath until the system reaches thermal equilibrium. Throughout this process, the volume of the gas remains constant (box length $L_{\textbf{b}}$ is unchanged), and the system remains in the TG regime, with no change in the interactions between particles. As a result of the heat exchange, the entropy of the system increases ($\Delta S > 0$) and the heat input is given by $\langle Q_2\rangle = \sum_{n}E^{(\textbf{b})}_{n} \left(f^{(\textbf{b})}_n(T_h)-f^{(\textbf{b})}_n(T_c') \right)$.
    
     \item \textbf{c $\rightarrow$ d} \textit{Adiabatic expansion}: The system is decoupled from the hot bath and the box is expanded from the compressed length $L_{\textbf{b}}$ back to the original length $L_{\textbf{a}}$. Simultaneously, the interactions between the particles are tuned from $c\rightarrow\infty$ to $c=0$ to return to the ideal Bose gas phase. During this stroke work is extracted from the system and is given by $ \langle W_3 \rangle = \sum_{n}\left( b^{(\textbf{a})}_n(T_h') E^{(\textbf{a})}_n - f_n^{(\textbf{b})}(T_h)E^{(\textbf{b})}_n \right)$ with $T_h'$ the effective temperature of the ideal Bose gas that ensures $\Delta S=0$.
     
     \item \textbf{d $\rightarrow$ a} \textit{Isochoric cooling}: In the final stroke of the \textbf{A}-cycle, the system is connected to a cold thermal reservoir at temperature $T_c<T_h$, allowing it to release heat and reach thermal equilibrium. During this stroke, the length of the box remains constant at $L_{\textbf{a}}$, and there is no change in the interaction strength as the system remains in the ideal Bose gas phase. The heat released during this stroke is given by $\langle Q_4\rangle =\sum_{n}E^{(\textbf{a})}_{n} \left[b^{(\textbf{a})}_n(T_c)-b^{(\textbf{a})}_n(T_h') \right].$
\end{enumerate}

The \textbf{A}-engine permits two different ways to change the statistics during engine operation. We term the cycle we have just described (and which is shown in Fig.~\ref{fig:InteractionEngine}(a)) the BG-TG cycle, as the initial state at point \textbf{a} is in the ideal bosonic gas phase and it is driven to the TG limit at point \textbf{b}. Schematically it can represented in a simplified entropy-inverse volume diagram as in Fig.~\ref{fig:InteractionEngine}(b), with the different statistical regimes highlighted by orange (BE statistics) and green (FD statistics). However, the order of the statistics change can also be reversed, as shown in Fig.~\ref{fig:InteractionEngine}(c), with the initial state at point \textbf{a} being in the TG limit and therefore described by the FD distribution, while at point \textbf{b} the system is in the ideal Bose gas phase described by the BE distribution. We therefore term this the TG-BG cycle, and will show in the following section that it displays a very different performance to the BG-TG cycle.  

\subsection{\textbf{A}-engine performance}

The two primary figures of merit for a heat engine are its work output and its efficiency, which are defined respectively as
 \begin{align}
     \mathcal{W}_\text{\tiny{A}} = - \left(\langle W_1\rangle + \langle W_3\rangle \right) && 
     \eta_\text{\tiny{A}} =  \mathcal{W}_\text{\tiny{A}}/\langle Q_2\rangle.
     \label{eq:otto_eff}
 \end{align}
The work output ($\mathcal{W}\text{\tiny{A}}$) is the net work produced by the engine over the entire cycle. For the system to function as an engine, the positive work condition must hold $\mathcal{W}\text{\tiny{A}} > 0$. This condition is achieved when the work extracted during the expansion stroke ($|W_3|$) exceeds the work invested during the compression stroke ($|W_1|$), i.e., $|W_3| > |W_1|$. The efficiency quantifies how much of the heat input $Q_2$ is transformed into usable work. Throughout the rest of the manuscript we will scale the work output in terms of the Fermi energy at point \textbf{b}, $E_F=N^2\pi^2/L_{\textbf{b}}^2$, and temperatures in units of Fermi temperature $T_F=E_F/k_B$.

The impact of particle statistics on the engine’s performance can be quantified by comparing to cycles which do not utilize a change in particle statistics. In particular we compare to: 
\begin{itemize}
    \item \textit{Single-particle engine}: The working medium consists of one particle with a thermal distribution described by $p_n=Z^{-1} e^{-E_n/T}$ with $Z=\sum_n e^{-E_n/T}$ the partition function. Such a  system is devoid of statistics or many-body interactions, and can serve as a baseline for comparison.
    
    \item \textit{Fully fermionic engine}: The working medium is composed entirely of fermionic particles and the engine operates under Fermi-Dirac statistics throughout the cycle.
    
    \item \textit{Fully bosonic engine}: The working medium is composed entirely of bosonic particles and the engine follows Bose-Einstein statistics throughout the cycle.
\end{itemize}
In all three cases the efficiency is given by the well known Otto efficiency $\eta_{\mathcal{O}}= 1- \left(L_\textbf{b}/L_\textbf{a} \right)^2$~\cite{Chen_2018}, while the work output can vary depending on the working medium. 

\begin{figure*}[tb]
     \centering
     
         \includegraphics[width=1\textwidth]{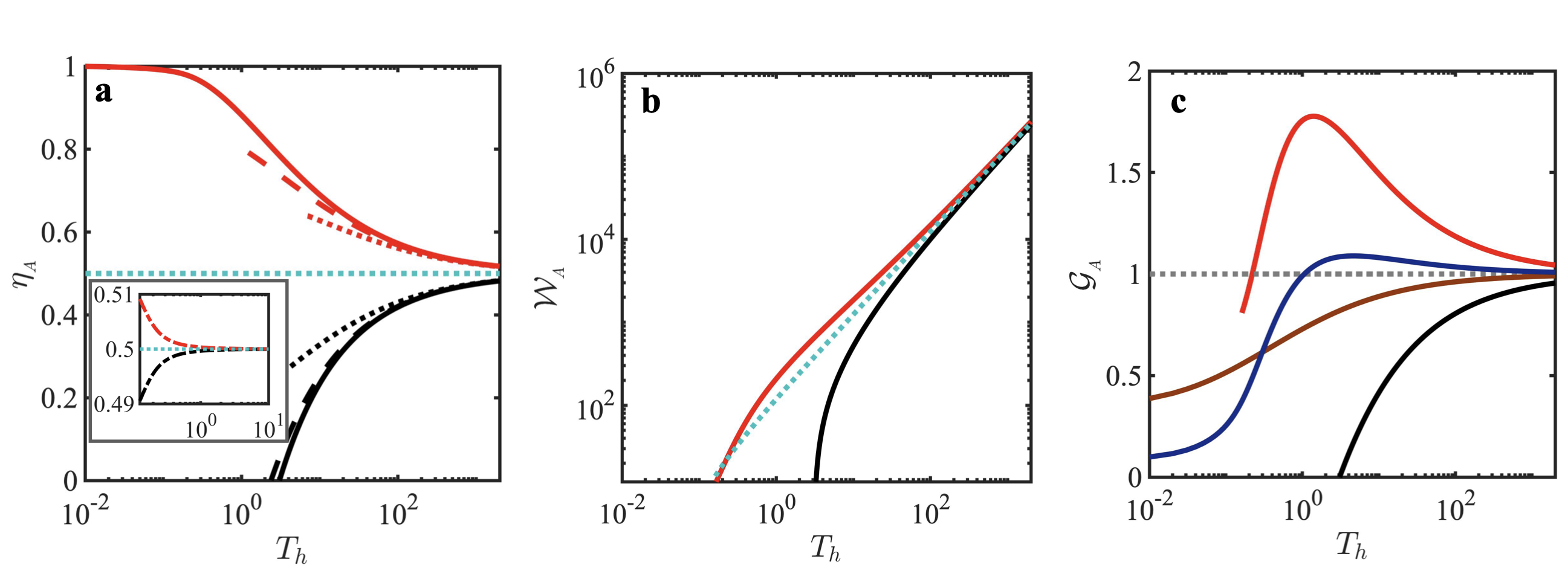}
     
        \caption{Performance of the \textbf{A}-engine as a function of the hot bath temperature $T_h$. (a) Efficiency of the BG-TG cycle (red) and the the TG-BG cycle (black) for cold bath temperatures $T_\text{c}=0$ (solid), $0.25$ (dashed) and $2.5$ (dotted). The Otto efficiency is shown for comparison as the dotted cyan line. The inset shows the case when the \textbf{A}-engine is executed in a 1D harmonic potential for a cold bath temperature of $T_c = 0$ and comrpession ratio $\omega_\textbf{b}/\omega_\textbf{a}=2$. (b) Work output of the \textbf{A}-engine scaled to the Fermi energy $E_F$ for the different cycles, with fixed cold bath temperature $T_c=0$. For comparison the work output of $N$ single-particle engines $N\mathcal{W}_1$ is shown by the cyan dotted line. (c) The work gain for the statistical engines, compared to the gain for a fully fermionic engine (blue) and a fully bosonic engine (brown).}
    \label{fig:PerformanceA}
\end{figure*}   

The analytic treatment of the LL model, including finite interaction strengths and finite temperatures, has been thoroughly discussed in \cite{kerr2024analytic} and allows some preliminary insight into the role statistics will in the performance of the two cycles of the \textbf{A}-engine.  
For simplicity, we consider the cold reservoir temperature to be $T_c = 0$ and the temperature of the hot bath such that the system stays in the degenerate regime $T_h < 1$. This ensures that the quantum statistical effects dominate and the efficiencies can then be calculated using the closed-form expressions for energy and entropy (see Appendix A for details).

Let us first focus on the BG-TG cycle where the efficiency can be expressed as
\begin{equation}
   \eta_\text{A}\sim 1 - \frac{8 \pi^3}{\left[9\sqrt{3} \times \zeta(3/2)\right]^2} T_h \left(\frac{L_\textbf{b}}{L_\textbf{a}}\right)^2 \,,
   \label{analyticalBG_TG_A}
\end{equation} 
 and which will only exceed the Otto efficiency, $\eta_{_A}>\eta_\mathcal{O}$, when the factor multiplied by the compression ratio $\left(\frac{L_\textbf{b}}{L_\textbf{a}}\right)^2$ is smaller than $1$. According to Eq.~\eqref{analyticalBG_TG_A} this requires $T_h \lesssim  6.68$, which is automatically satisfied in the degenerate regime we are considering. This result as a function of the hot bath temperature $T_h$ is confirmed by our numerical calculations shown in Fig.~\ref{fig:PerformanceA}(a) for system of $N=500$ bosons and a fixed compression ratio $\left(L_\textbf{b}/L_\textbf{a}\right)^2=1/2$. The efficiency of the BG-TG cycle is indicated by the red lines (different line styles represent different temperatures of the cold bath $T_c$), while the Otto efficiency $\eta_{\mathcal{O}}=1/2$ is indicated by the dotted cyan line. Immediately it is apparent that the efficiency of the BG-TG cycle greatly exceeds the Otto efficiency for the entire temperature range considered, which is an impressive demonstration of how the change of statistics enhances the performance of the engine. By changing the statistics during the compression and expansion strokes the work done is increased, which results in the improved performance. The increase is most notable at low temperatures where effects of the statistics is dominant, with the efficiency approaching $1$ as $T_h\rightarrow 0$.

In comparison, for the TG-BG cycle the efficiency is found to be
\begin{equation}
   \eta_\text{A}\sim 1 - \frac{ (3\sqrt{3})^2\zeta(3/2)}{4\pi\sqrt{\pi}} T_h^{-1/2}\left(\frac{L_\textbf{b}}{L_\textbf{a}}\right)^2\,.
   \label{analyticalTG_BG_A}
\end{equation}  
Unlike the BG-TG cycle, the efficiency is proportional to $T_h^{-1/2}$, which significantly impacts its performance. As the Otto efficiency can only be exceeded if $T_h\gtrsim 10$, which is far outside the regime of validity of this expression ($T_h<1$), Eq.~\eqref{analyticalTG_BG_A} will always give $\eta_A<\eta_{\mathcal{O}}$.  This is confirmed by the numerical results shown as black lines in Fig.~\ref{fig:PerformanceA}(a), where one can see the TG-BG cycle consistently underperforms and even has vanishing efficiency for $T_h<4$. It therefore does not operate as an engine in this regime. The reason for the reduced efficiency is due to competition between the work done due to the change in interaction and that being done by changing the trap length. They possess opposite signs when implemented during the same stroke, which reduces the amount of work done during the compression, but more crucially reduces the amount of work extracted during the expansion stroke.

One can also see that the statistical effects are only nullified once the system has reached significantly hot bath temperatures, which leads to the BG-TG and TG-BG cycles asymptotically approaching the Otto efficiency (see also the analytic expressions in Appendix A). However, for the BG-TG cycle the efficiency also reduces when increasing the temperature of the cold bath (dashed and dotted red lines), which restricts the operational regime of the engine. The robustness of quantum statistical effects between the ideal Bose gas and the TG gas is due to the quadratic single particle energy spectrum of the box potential and its decreasing density of states with increasing energy, $g_{\text{1D}}(E) \propto E^{-1/2}$ \cite{kerr2024analytic}. Thermodynamically this means that the cost for excitations $\langle Q_2 \rangle$ is very high, and therefore statistical effects are preserved even at large temperatures. In comparison, statistical effects are washed out at much lower temperatures ($T_h\gtrsim1$) in the harmonic oscillator (see inset of Fig.~\ref{fig:PerformanceA}(a)) as the constant density of states $g_{\text{1D}}(E) \propto E^{0}$ is more conducive to thermal excitations \cite{Zheng2015}. In any case, as the temperature of the system approaches the classical regime, the efficiencies of both cycles reduce to the Otto efficiency owing to the fact that distinct quantum statistics of either phase is no longer available.

In Fig.~\ref{fig:PerformanceA}(b) we show the work output versus $T_h$ for both the BG-TG and TG-BG cycles with $T_c=0$, and compare this to the work output of $N$ single particle heat engines (dotted cyan line). One can see that the BG-TG cycle has the largest work output, superseding that of the single particle engines, particularly around the Fermi temperature $T_h=1$. This is more apparent in Fig.~\ref{fig:PerformanceA}(c) where we plot the ratio of work output compared to the $N$ single particle engines, $\mathcal{G}= \mathcal{W}^{(N)}/(N \mathcal{W}_1)$. The BG-TG cycle has a maximum performance gain of $\mathcal{G}\approx 1.78$ and easily outperforms purely bosonic (brown line) and fermionic (blue line) engines. This boost to the work output is a combination of the statistics change significantly increasing the amount of work done during both the compression and expansion strokes, while also affecting the amount of heat that is exchanged between the two baths. For instance, the heat input is equivalent to the purely fermionic engine, $\langle Q_2 \rangle=\langle Q^{\textbf{F}}_2 \rangle$, however the amount of heat dumped into the cold bath is reduced as the BG-TG engine is in the ideal bosonic gas phase during the isochoric cooling stroke, ensuring that $|\langle Q_4 \rangle|<|\langle Q^{\textbf{F}}_4 \rangle|$. It is this difference in heat energy that has been converted into usable work in the BG-TG cycle, $\mathcal{W}_A-\mathcal{W}_F=|\langle Q^{\textbf{F}}_4 \rangle|-|\langle Q_4 \rangle|$, and is responsible for the increased work output. At high temperatures all quantum statistical effects are lost and the work output converges to the single particle engine result as also seen in the efficiency.

\begin{figure*}[tb]
    \centering
     \includegraphics[width=1\linewidth]{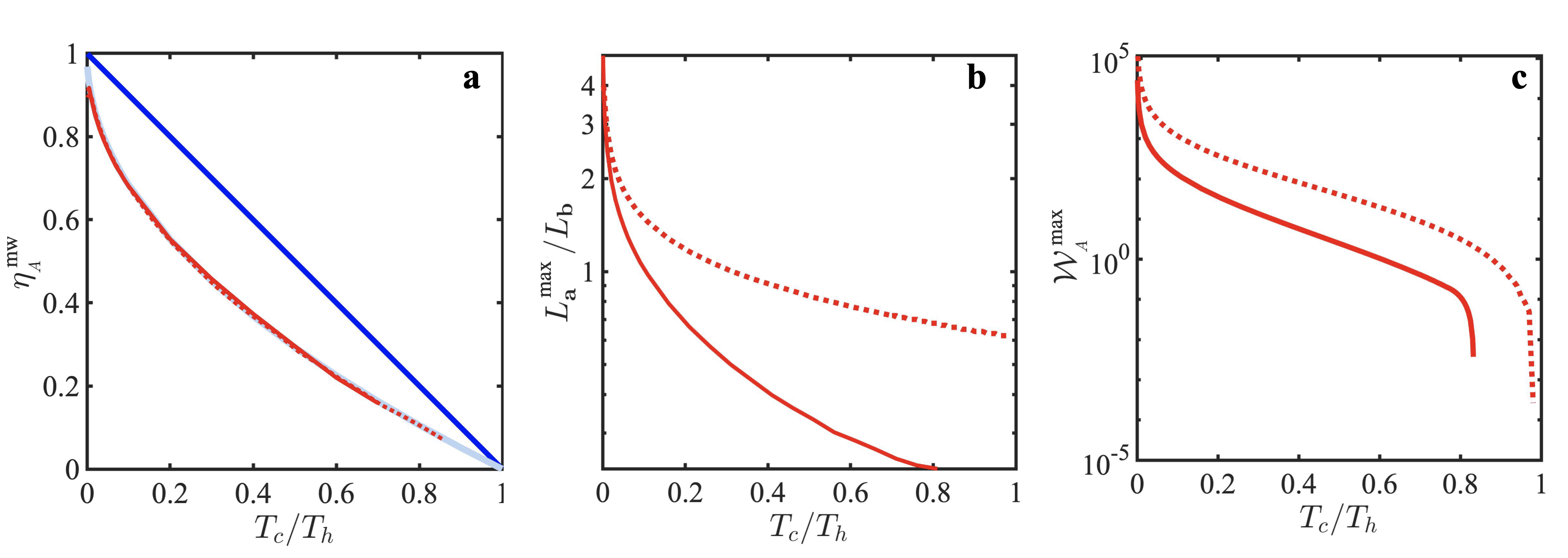}
   \caption{ (a) Efficiency at maximum work, $\eta^\text{mw}_{_A}$ as a function of $T_{c}/T_{h}$, for two different temperatures of the cold bath, $T_{c}= 0.1$ (solid red line) and  $T_{c}= 1 $ (dotted red line). For comparison, the Carnot efficiency (dark blue solid line) and the Curzon-Ahlborn (CA) efficiency (light blue solid line) is shown. (b) The optimal length ratio, $L^\text{max}_{\textbf{a}}/L_{\textbf{b}}$, and (c) the maximum work output scaled to the Fermi energy, $\mathcal{W}^{\text{max}}_A$, for the same values of the cold bath temperature as panel (a).
   } 
    \label{fig:Optimization_A}
\end{figure*}

Finally, we explore the efficiency at maximum work output for the \textbf{A}-engine with the  BG-TG cycle, $\eta^\text{mw}_{_A}$, by optimizing the compression ratio $\left(L_\textbf{b}/L_\textbf{a}\right)^2$. We show $\eta^\text{mw}_{_A}$ in Fig.~\ref{fig:Optimization_A}(a) as a function of $T_c/T_h$ for two fixed cold bath temperatures, $T_c=0.1$ (solid line) and $T_c=1$ (dotted lines). The upper bound on the efficiency for any heat engine is given by the Carnot efficiency $ \eta_\text{\tiny{C}}= 1 - T_c/T_h$ (blue solid line) which is an immediate consequence of the second law. While one can see that $\eta^\text{mw}_{_A}$ lies below the Carnot efficiency we find that it actually follows the Curzon-Ahlborn (CA) bound (light blue solid line), $\eta_\text{\tiny{CA}} = 1 - \sqrt{T_c/T_h }$~\cite{Curzon_1975,Leff_1987}, which has been shown to hold for the quantum Otto cycle at high temperatures and is a good indicator of the performance of our engine~\cite{Rezek_2006}. That $\eta^\text{mw}_{_A}$ is given by the CA bound should not be surprising, since the change in statistics can be seen as simply enhancing the work done during the adiabatic strokes and therefore the \textbf{A}-engine is essentially a modified Otto engine. However, it is worth noting that $\eta^\text{mw}_{_A}$ is not affected by the cold bath temperature $T_c$, indicating that our engine cycle operates at the optimal efficiency regardless of being in, or out of, the quantum degenerate regime. While a similar effect has been seen in few-body interacting heat engines when the cold bath is in the degenerate regime $T_c\ll1$ \cite{Mohammed_2023}, the CA bound is only reached when the hot bath has similarly low temperatures. 

On the other hand, the cold bath temperature does affect the operational range of the \textbf{A}-cycle, as shown in Fig.~\ref{fig:Optimization_A}(a), which places a limit on the maximum $T_c/T_h$ that is accessible (note that for $T_c/T_h\rightarrow1$ there is no data in Fig.~\ref{fig:Optimization_A}(a)). This can be understood by examining the optimal box length at point \textbf{a} that gives the maximum work output, $L^\text{max}_{\textbf{a}}/L_{\textbf{b}}$, shown in Fig.~\ref{fig:Optimization_A}(b), and the maximum work output shown in Fig.~\ref{fig:Optimization_A}(c). For large hot bath temperatures, $T_c/T_h\rightarrow0$, the length of the box trap is reduced during the $\langle W_1 \rangle$ work stroke, $L^\text{max}_{\textbf{a}}>L_{\textbf{b}}$, compressing the gas and decreasing its volume as expected. However, for small hot bath temperatures, $T_c/T_h\rightarrow1$, we find that the optimal box length at point \textbf{b} is larger, $L^\text{max}_{\textbf{a}}<L_{\textbf{b}}$, which corresponds to an increase of the trap volume and therefore the gas expands during the $\langle W_1 \rangle$ work stroke. While this contradicts the usual stroke order of the Otto cycle, the cycle also does work by changing the statistics. If the the change in interaction energy dominates over the change in volume the signs of the work strokes can be preserved, $\langle W_1 \rangle>0$ and $\langle W_3 \rangle<0$, and the cycle still operates as an engine in this low temperature regime. However, this adds a limitation in terms of the lowest operational temperature for the engine with the work output vanishing for finite $T_h>T_c$ as shown in Fig.~\ref{fig:Optimization_A}(c). In this case, the change in interaction energy cancels with the change in volume and no more work can be extracted from the cycle.

\subsection{Statistically enhanced thermal strokes: T-engine}

\begin{figure*}[tb]
    \centering
    \includegraphics[width=0.85\linewidth]{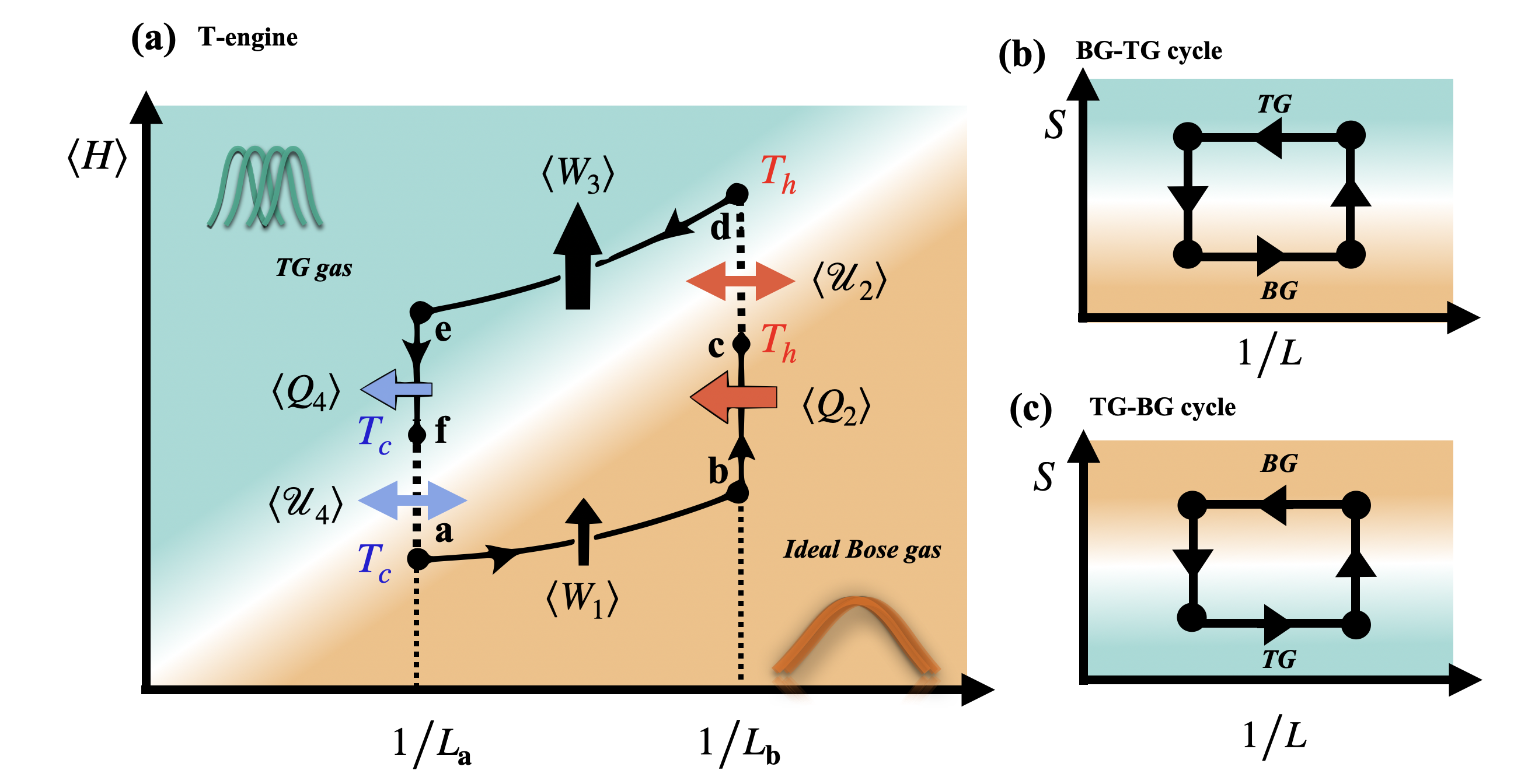}
    \caption[Schematic of the \textbf{T}-engine]{(a) Schematic of the \textbf{T}-engine with the BG-TG cycle, where the particle statistics is changed during the isothermal strokes \textbf{c} $\to$ \textbf{d} and \textbf{f} $\to$ \textbf{a}, where the total energy change is given by $\langle \mathscr{U} \rangle$. During the adiabatic strokes the particle statistics are fixed to being bosonic during compression and fermionic during expansion, leading to work $\langle W_1\rangle$ and $\langle W_3\rangle$, respectively. (b) Entropy-inverse volume diagram for the BG-TG and (c) for the TG-BG cycles of the \textbf{T}-engine.}
    \label{fig:T-cycleScheme}
\end{figure*}

Next we discuss an alternative Otto-like cycle in which the statistics are changed during the isochoric thermal strokes, and therefore the adiabatic compression and expansion strokes are implemented in the different statistical regimes. We term this the \textbf{T}-engine and show the corresponding BG-TG cycle schematically in Fig.~\ref{fig:T-cycleScheme}(a). Here, the compression stroke is carried out in the ideal Bose gas regime, while the expansion stroke occurs in the Tonks--Girardeau regime. 
The individual strokes are carried out as follows: 
\noindent
\begin{enumerate}

    \item $\textbf{a} \to \textbf{b}$ \textit{Adiabatic compression}:  At point \textbf{a} the system is in the ideal Bose gas phase and thus described by the Bose-Einstein distribution with temperature $T_c$. The interaction strength is $c=0$ and therefore the single particle statistics is unchanged, while the length of the box is adiabatically ($\Delta S=0$) compressed from $L_{\textbf{a}}$ to $L_{\textbf{b}}$ which does work 
    $ \langle W_1\rangle = \sum_{ n} b^{(\textbf{a})}_{n}(T_c)\left[E^{(\textbf{b})}_{n}-E^{(\textbf{a})}_{n}\right]$.

    \item $\textbf{b} \to \textbf{c}$ \textit{Isochoric heating}: At fixed length $L_{\textbf{b}}$ the system is brought into contact with a hot bath at temperature $T_h$ and heat is exchanged until the system achieves equilibrium. During this process the system remains in the Bose gas phase with no change in the interaction and therefore the heat input is given by $\langle Q_2\rangle = \sum_{n}E^{(\textbf{b})}_{ n} \left[b^{(\textbf{b})}_{n}(T_h)-b^{(\textbf{a})}_{n}(T_c) \right]$. 

    \item $\textbf{c} \to \textbf{d}$ \textit{Statistical hot isotherm}: While still coupled to the hot bath the system is driven to the Tonks--Girardeau regime ($c \rightarrow \infty$) and the statistics changes from Bose-Einstein to Fermi-Dirac, with the total energy change during this stroke given by $\langle \mathscr{U}_2\rangle = \sum_{n} E^{(\textbf{b})}_n\left[ f^{(\textbf{b})}_{n}(T_h) - b^{(\textbf{b})}_{n}(T_h) \right]$. However, this stroke contains contributions from both heat and work which must be accounted for separately, i.e. $\langle \mathscr{U}_2\rangle=\langle Q_2\rangle^{s}+\langle W_2\rangle^{s}$. The heat contribution can be calculated from the change in entropy from the fermionic to bosonic statistics at temperature $T_h$, $\langle Q_2\rangle^{s}=\left( S_F^{(\textbf{b})}(T_h)-S_B^{(\textbf{b})}(T_h) \right)T_h$ and the work done by the change in statistics can therefore be calculated from $\langle W_2 \rangle^s = \langle \mathscr{U}_2\rangle-\langle Q_2\rangle^{s}$.

    \item $\textbf{d} \to \textbf{e}$ Adiabatic expansion: The system is decoupled from the bath and the box is adiabatically expanded back to its original length from $L_\textbf{b}$ to $L_\textbf{a}$. Throughout the expansion stroke the system remains in the TG gas regime with a fixed interaction strength, and described by the Fermi-Dirac distribution. The extracted work is given by $ \langle W_3\rangle = \sum_{n} f^{(\textbf{b})}_{n}(T_h)\left[E^{(\textbf{a})}_{n}-E^{(\textbf{b})}_{n}\right]$.

    \item $\textbf{e} \to \textbf{f}$ Isochoric cooling: The system is coupled to a cold bath at temperature $T_c$ at fixed box length $L_{\textbf{a}}$ and interaction $c\rightarrow \infty$. Heat is exchanged with the cold bath until the system reaches equilibrium, with the heat released given by $\langle Q_4\rangle =\sum_{n}E^{(\textbf{a})}_{n} \left[f^{(\textbf{a})}_{n}(T_c)-f^{(\textbf{b})}_{n} (T_h) \right]$.  

    \item $\textbf{f} \to \textbf{a}$ Statistical cold isotherm: The interactions are tuned from $c\rightarrow\infty$ (fermionic statistics) to $c=0$ (bosonic statistics) while coupled to the cold bath at temperature $T_c$. Again, the change in internal energy has two distinct contributions $\langle \mathscr{U}_4\rangle = \langle W_4\rangle^{s} + \langle Q_4\rangle^{s}$, where $\langle \mathscr{U}_4\rangle = \sum_{n} E^{(\textbf{a})}_n\left[ b^{(\textbf{a})}_{n}(T_c)-f^{(\textbf{a})}_{n}(T_c) \right]$ and $\langle Q_4\rangle^{s} = \left(S_B^{(\textbf{a})}(T_c) - S_F^{(\textbf{a})}(T_c)\right)T_c$.
    
\end{enumerate}

Similarly, the TG-BG variant of the \textbf{T}-engine can be realized by performing the compression stroke in the TG gas regime and the expansion stroke in the ideal Bose gas regime. 

For \textbf{T}-engine, the work output is modified to include the extra work terms coming from the statistical isotherms
\begin{equation}
    \mathcal{W}_\text{\tiny{T}} = -\left(\langle W_1\rangle + \langle W_2\rangle^{s} + \langle W_3\rangle +\langle W_4\rangle^{s}\right),
\end{equation} while the efficiency includes the heat change due to the change in statistics during the hot isotherm
\begin{equation}
    \eta_\text{\tiny{T}} = \frac{ \mathcal{W}_\text{\tiny{T}} }{\left(\langle Q_2\rangle +\langle Q_2\rangle^{s}\right)}\,.
\end{equation}
These extra contributions to the different strokes will have a significant effect on the performance of the \textbf{T}-engine, which we will explore in the next section.

\subsection{\textbf{T}-engine performance}
We can approximate the work output and efficiency for both the BG-TG and TG-BG cycles using the analytic expressions for the energy and entropy, as done before for the \textbf{A}-engine. 
For simplicity we assume $T_c=0$ and therefore the last stroke only has a contribution from the work done during the change in statistics between the ground state of the ideal Bose gas and the TG gas, $\langle \mathscr{U} \rangle_4=\langle W_4\rangle^s$.

\begin{figure*}[tb]
        \includegraphics[width=1\linewidth]{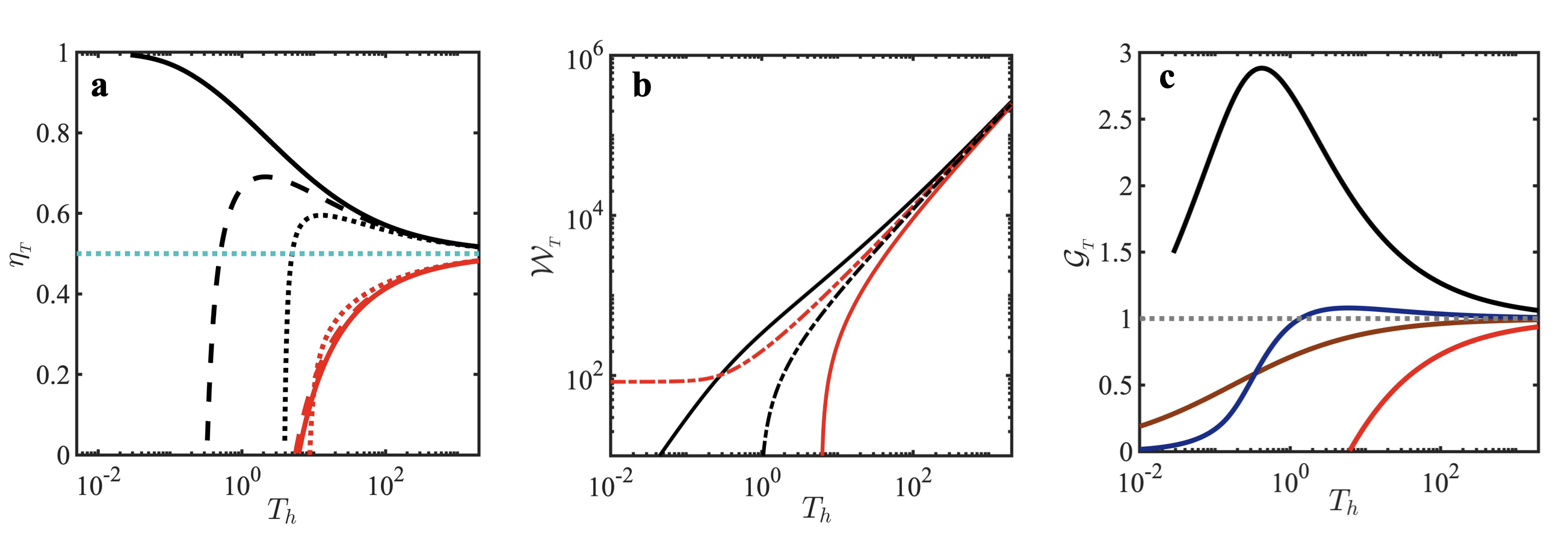}
   \caption[Performance of the T-engine]{Performance of the \textbf{T}-engine as a function of temperature of the hot bath $T_h$. Panel (a) shows the efficiency for the BG-TG cycle in red and for the TG-BG cycle in black, and for the cold bath temperatures of $T_\text{c}=0$ (solid), $T_\text{c}=0.25$ (dashed) and $T_\text{c}=2.5$ (dotted). The dashed cyan line indicates the Otto efficiency. Panel (b) shows the work output for the \textbf{T}-engine (solid lines)  and for the \textbf{GV}-engine (dash-dotted lines) for $T_\text{c}=0$, with the above color coding to identify the BG-TG (red) and TG-BG (black) cycles. In panel (c) the work output gain is shown in comparison with the gain for the fully fermionic engine (blue line) and the fully bosonic engine (brown line).}
    \label{fig:T-performace}
\end{figure*}

For the BG-TG engine, the work output in units of Fermi energy is given by
\begin{equation}
    \mathcal{W}_\text{T} \sim N \left[ \frac{T_h^2}{6} - \frac{\zeta(3/2) \sqrt{\pi}}{2} T_h^{3/2} - \pi^2\frac{T_h^2}{12}\right] \left( \frac{L_b}{L_a}\right)^2 
\end{equation}
which is negative for temperatures $T_h<1$ and therefore this cycle does not operate as an engine in the degenerate regime. This is confirmed by the numerical calculations shown in Fig.~\ref{fig:T-performace} and one can see that the BG-TG cycle requires hot bath temperatures on the order of several Fermi temperatures to fulfill the positive work condition. For the TG-BG cycle, the efficiency in the low temperature limit is given by
\begin{equation}
        \eta_\text{T} \sim 
        1- \left[\frac{3 \pi^3\zeta(3/2)T_h^{3/2}}{9\zeta(3/2)-\pi^{9/2}T_h^2} \right] \left( \frac{L_b}{L_a}\right)^2
\end{equation}
which is larger than the Otto efficiency in the degenerate regime $T_h\ll1$. As shown in Fig.~\ref{fig:T-performace}(a), the TG-BG cycle is capable of exceeding the Otto efficiency for a large range of hot bath temperatures when $T_c=0$ (solid black line). However, finite cold bath temperatures reduce the efficiency (dotted and dashed black lines), with $\eta<\eta_O$ for $T_h \approx T_c$. In comparison, the efficiency of BG-TG cycle is practically unchanged when increasing cold bath temperature. Furthermore, in Fig.~\ref{fig:T-performace}(b) the work output of the TG-BG cycle significantly exceeds the BG-TG cycle, and possesses nearly a three-fold gain in work output compared to $N$ single particle engines as shown in Fig.~\ref{fig:T-performace}(c).

The performance of the \textbf{T}-engine can therefore significantly exceed that of the \textbf{A}-engine, however, in what may seem to be a counter-intuitive manner by changing the statistics in the opposite order: going from fermionic to bosonic statistics after the initial compression stroke. While this reduces the internal energy of the system by removing the interactions between particles, it also crucially alters the entropy of the state. In a 1D box potential at temperature $T_h$ the entropy of the ideal Bose gas is larger than that of the TG gas, $S_B(T_h)>S_F(T_h)$, primarily owing to the lower density of states at the Fermi energy compared to bottom of the energy spectrum. Changing from fermionic to bosonic statistics while connected to the hot bath therefore results in a positive change in entropy and consequently there is additional heat flow into the system $\langle Q_2\rangle^s>0$. The engine operating with the TG-BG cycle can then absorb more heat from the hot bath which can be converted into more usable work, while the opposite effect is seen with the BG-TG cycle where $\langle Q_2\rangle^s<0$ resulting in diminished performance. In the TG-BG cycle work is also extracted during this stroke $\langle W_2\rangle^s <0$ due to the aforementioned reduction of interaction energy which also adds to the total amount of work outputted by the engine.

\begin{figure*}[tb]
    \centering
    \includegraphics[width=1\linewidth]{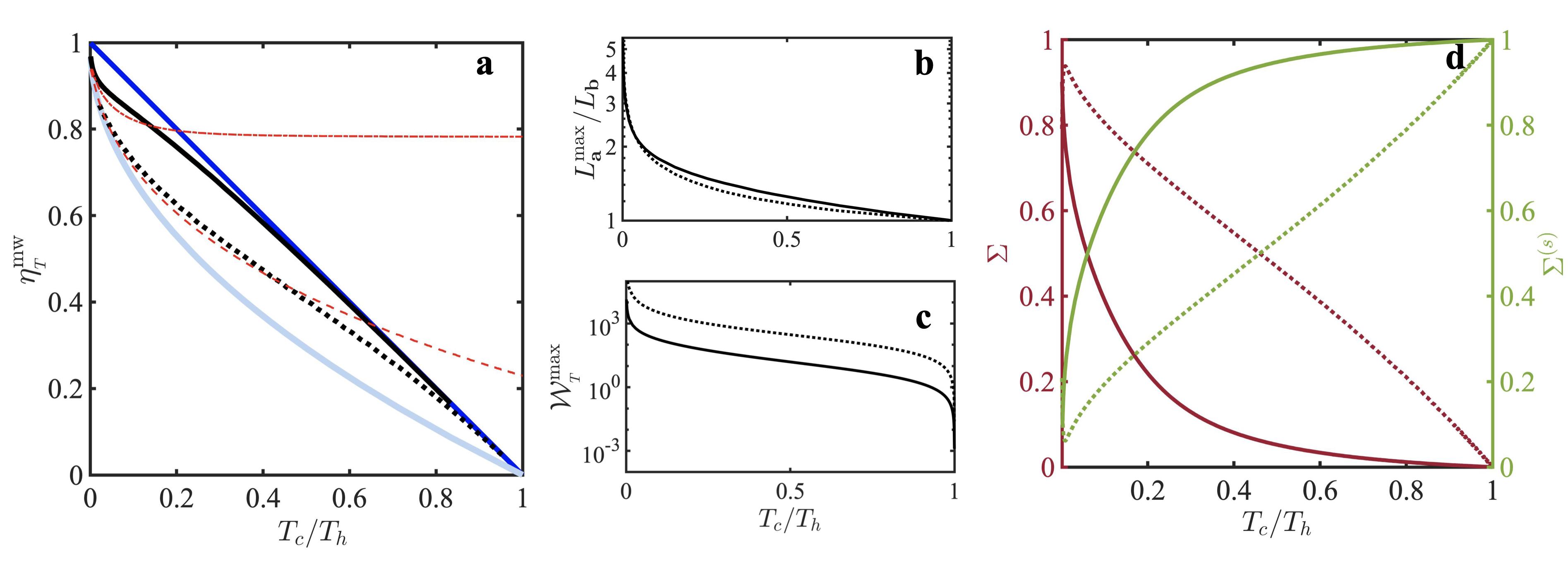}
   \caption{(a) The efficiency at maximum work, $\eta^\text{mw}_{_T}$, for the \textbf{T}-engine TG-BG cycle plotted as a function of $T_{c}/T_{h}$. Here the black solid curve and dotted curve are for a cold bath temperatures $T_{c}= 0.05 $ and  $T_{c}= 2$ respectively, with the dark blue curve showing the Carnot bound and the light blue curve showing the CA bound. The red curves show the optimized efficiency of the \textbf{GV}-engine BG-TG cycle for $T_c = 0.05$ (dashed-dotted) and $T_c = 2$ (dashed).
   (b) The optimal length ratio, $L^\text{max}_{\textbf{a}}/L_{\textbf{b}}$ and (c) the maximum work output in units of the Fermi energy for \textbf{T}-engine TG-BG cycle. (d) The ratio of thermal, $\Sigma$ (maroon curves), and statistical $\Sigma^{(s)}$ (green curves), heating  are shown for $T_c = 0.05$ (solid lines) and $T_c = 2$ (dotted lines).}
    \label{fig:Optimization_T}
\end{figure*}

In Fig.~\ref{fig:Optimization_T}(a) we show the efficiency at maximum work output, $\eta^\text{mw}_{_T}$ for the \textbf{T}-engine with the TG-BG cycle and for two representative cold bath temperatures, $T_c=0.05$ (solid black curve) and $T_c=2$ (dotted black curve). In Fig.~\ref{fig:Optimization_T}(b) and (c) we show the optimal trap length and maximum work output respectively. One can see immediately that $\eta^\text{mw}_{_T}$ exceeds the CA bound (light blue solid line) over the entire range of $T_h$, and indeed can saturate to the Carnot bound when both $T_c$ and $T_h$ are small, i.e.~within the degenerate regime where the changes in quantum statistics are dominant. The increase in efficiency is significant when compared to the \textbf{A}-engine in Fig.~\ref{fig:Optimization_A}(a), where the CA bound is never exceeded, showing that the interplay between work and heat during the isochoric strokes of the \textbf{T}-engine allows for increased performance.

To assess how the change in statistics affects the heat input we calculate the ratio of heat absorbed during the isochoric heating stroke (\textbf{b}$\rightarrow$\textbf{c}) and that from the statistical hot isotherm (\textbf{c}$\rightarrow$\textbf{d}), given respectively by 
\begin{align}
    \Sigma = \frac{\langle Q_2\rangle}{\left(\langle Q_2\rangle + \langle Q_2\rangle ^{s} \right)}, && 
    \Sigma^{(s)} = \frac{\langle Q_2\rangle^{s}}{\left(\langle Q_2\rangle + \langle Q_2\rangle ^{s} \right)}\,.
    \label{Qratio}
\end{align}
These are shown in Fig.~\ref{fig:Optimization_T}(d) for $T_c=0.05$ (solid) and $T_c=2$ (dashed). The statistical heat contribution clearly dominates at low hot bath temperatures, $T_c/T_h\rightarrow1$, where the Carnot bound is reached, while it vanishes at large hot bath temperatures, $T_c/T_h\rightarrow0$, where $\eta^\text{mw}_{_T}$  collapses to the CA bound as expected in the classical regime. The importance of the statistical heat in the performance of the \textbf{T}-engine is clear in the degenerate regime ($T_c=0.05$) where it constitutes the majority of the heat input for a large range of $T_h$, and is responsible for the engines ability to operate at the Carnot efficiency. 

\subsubsection{Global Variable Otto engine}
We emphasize that the statistical boost observed in the \textbf{T}-engine relies on separating the heat and work contributions in the isothermal strokes. To stress this point we can compare to engine cycles in which we consider that the energy differences during the thermal strokes, from point \textbf{b} to \textbf{d} and from point \textbf{e} to \textbf{a}, are wholly considered as heat and do not contribute to the total work of the engine.
This approach is consistent with the global thermodynamic variables description \cite{Rochin_2005, Rochìn_2008} which has been used to describe quantum engines in recent experiments~\cite{Koch2023,Simmons23,Ayala_2023}, and which recasts heat and work in terms of a global pressure and volume. In this approach, changes in gas volume, and thus work, can only be associated to changes in the trapping potential, which is now described by a trap volume ($\omega^{-1}$ for harmonic oscillator, and $L$ for the box), while isochoric processes are associated with changes in pressure at fixed trap volume.

Within this description the global variable Otto engine (\textbf{GV}-engine) will have the usual two adiabatic and two isochoric strokes, with efficiency and work output given by the standard expressions in Eq.~\eqref{eq:otto_eff}. The efficiency of the \textbf{GV}-cycle will be the Otto efficiency, regardless of the order of the statistics change, TG-BG or BG-TG. However, we do observe a significant difference in the work output, as shown by the dot-dashed lines in Fig.~\ref{fig:T-performace}(b). Interestingly, the BG-TG cycle exceeds the work output of the TG-BG cycle, as the work cost in changing from bosonic to fermionic statistics is now included with the heat input. The BG-TG cycle is a thermal variant of the Pauli engine \cite{Koch2023}, with the increased Fermi pressure leading to more work output. In Fig.~\ref{fig:Optimization_T}(a) we also show that the efficiency at maximum work, $\eta^\text{mw}$, of this cycle (red dashed and dot-dashed lines) can seemingly exceed the Carnot bound.

However, two important points need to be noted: firstly, for vanishingly small bath temperatures ($T_c\rightarrow0$ and $T_h\rightarrow0$) the work output and $\eta^\text{mw}$ of the \textbf{GV}-engine both saturate to constant values. In this regime the \textbf{GV}-engine does not operate as a quantum heat engine, as there is no heat input from the baths, instead it converts interaction energy to trap potential energy as in recent experiments with unitary quantum engines \cite{Koch2023,Simmons23}. The Carnot bound is then irrelevant as the engine is wholly fueled by a non-thermal energy source \cite{Niedenzu2018}. Secondly, when operating at finite temperatures ($T_h>0.3$) the work output of the TG-BG \textbf{T}-engine exceeds the \textbf{GV}-engine, as including the work contributions from the interaction ramps adds to the total work output of the heat engine cycle (see Fig.~\ref{fig:T-performace}(b)). This highlights the full power of this hybrid Otto engine and the importance of separating the distinct energy contributions coming from thermal baths and interactions in cold atom systems.

\subsection{Other thermal operations}

\begin{figure*}[t!]
    \centering
    \includegraphics[width=1\linewidth]{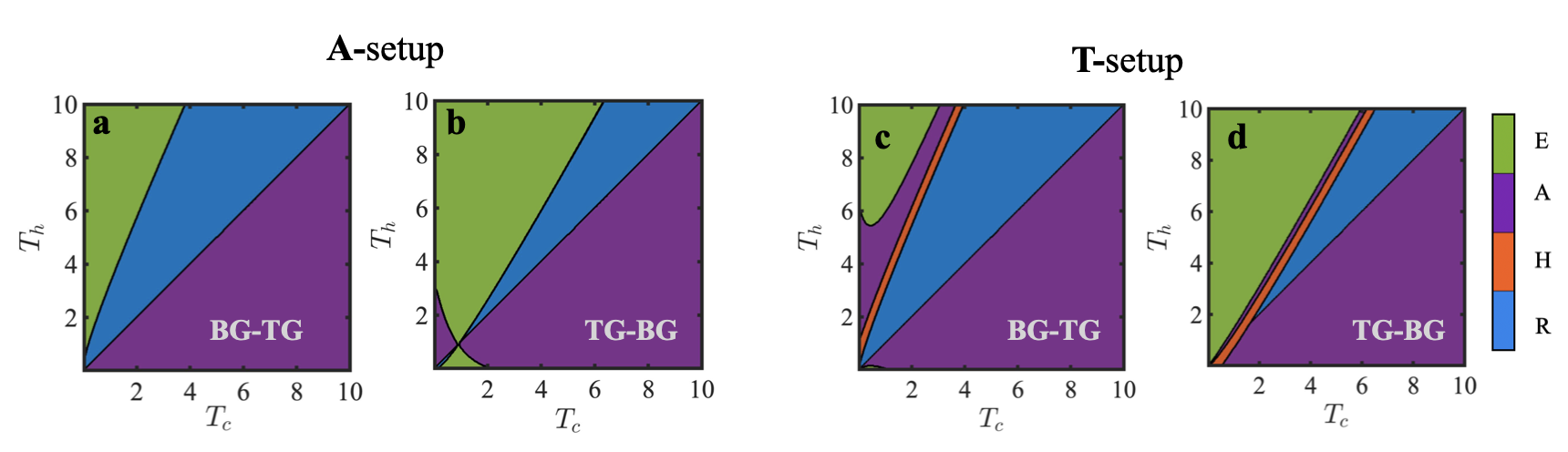}
   \caption{Thermal operational modes of the \textbf{A}-setup and the \textbf{T}-setup, as a function of the cold and hot bath temperatures. The different operational regimes are an engine (E - green), a refrigerator (R - blue), a heater (H - red) and an accelerator (A - purple). Panels (a) and (b) show the thermal operations for the BG-TG and TG-BG cycles in the \textbf{A}-setup, respectively, whereas panels (c) and (d) are the BG-TG and TG-BG cycles, respectively, for the \textbf{T}-setup. }
   \label{ThermalOP}
\end{figure*}
Finally, we discuss the possibility of the setup for the  \textbf{A}-engine and for the \textbf{T}-engine to be used to create other quantum thermal machines, namely, refrigerators, heaters and accelerators. The operational regimes of these different devices can be categorized 
through the total work, heat input and output in the following way: 
\noindent
\begin{enumerate} [itemsep=1mm] 
 
  \item  Engine \textbf{(E)}  \ \ \ \ \ \ \ :  $Q_\text{in} \geq 0$ \ $Q_\text{out} \leq 0$ \ $\mathcal{W} \geq 0$ 
   \item Refrigerator \textbf{(R)}\ : $Q_\text{in} \leq 0$ \ $Q_\text{out} \geq 0$ \ $\mathcal{W} \leq 0$ 
   \item Accelerator \textbf{(A)} \ : $Q_\text{in} \geq 0$ \ $Q_\text{out} \leq 0$ \ $\mathcal{W} \leq 0$ 
   \item Heater \textbf{(H)} \ \ \ \ \ \ \ :  $Q_\text{in} \leq 0$ \ $Q_\text{out} \leq 0$ \ $\mathcal{W} \leq 0$ 
   
\end{enumerate}
For the engine, as discussed above, heat input from the hot bath is converted into work that can be extracted $\mathcal{W}>0$. And when all these constraints are reversed, the cycle performs as a refrigerator, as work is needed to be added to further cool the cold bath. In a similar manner the accelerator uses work done on the system to accelerate heat flow from hot to cold baths. Finally, there is the heater where the work $\mathcal{W}$ done on the system is used to heat up both the cold and hot bath. 

We show the regimes for the different operational modes in Fig.~\ref{ThermalOP}. 
Let us first discuss the \textbf{A}-setup which is shown in Fig.~\ref{ThermalOP}(a) for the BG-TG cycle and Fig.~\ref{ThermalOP}(b) for the TG-BG cycle as a function of cold and hot bath temperatures. In both cases the system may operate as an engine, a refrigerator or an accelerator. However, operating as a heater is not possible as $Q_\text{in}$ and $Q_\text{out}$ cannot simultaneously exchange heat in the same manner. For the BG-TG cycle (Fig.\ref{ThermalOP}(a)) the engine and refrigerator operations are only found for the case when $T_c<T_h$ whereas the cycle operates as an accelerator for $T_c>T_h$, where the conditions for heat exchange are reversed. A more interesting case is that of the TG-BG cycle (Fig.~\ref{ThermalOP}(b)), where there is a difference in sign between the work done during the change in interaction energy and that done during the change of the box length. For low bath temperatures these two processes counteract each other leading to $\mathcal{W}<0$ and therefore the cycle operates as an accelerator even though $T_h>T_c$. A finite $T_h$ is then required to excite the ideal Bose gas to high enough energies to produce positive work output $\mathcal W>0$. This same interplay between the opposing work contributions from interactions and trap changes allows an engine to operate in a small parameter regime when $T_c>T_h$.  

The parameter regimes for the different operational modes for the \textbf{T}-setup are shown in Fig.~\ref{ThermalOP}(c) for the BG-TG cycle and in Fig.~\ref{ThermalOP}(d) for the TG-BG cycle. One can see that in the $T_h>T_c$ region of both cycles all modes of operation are possible: engine, accelerator, heater and refrigerator. Most notably a heater can be realized in the \textbf{T}-setup due to the interplay between the statistical work $\langle W\rangle^s$ and heat $\langle Q\rangle^s$ during the isothermal strokes which can used to heat up both baths. In the BG-TG cycle an engine can only be operated above a certain $T_h$, i.e.~once it is possible to excite the TG gas.
In comparison, for the TG-BG cycle, the positive work condition is readily met for a large parameter range in the region $T_h>T_c$ where it operates as an engine. In particular this shows that operating the TG-BG \textbf{T}-cycle heat engine is possible for larger cold bath temperatures than the BG-TG \textbf{A}-cycle, showing another advantage for this hybrid Otto scheme.  

\section{Conclusion}
We have analyzed two distinct types of Otto-like quantum heat engines, realized in lower dimensions and with a working medium made from ultracold atoms. The engines considered  used either statistically enhanced adiabatic or thermal strokes, thereby realizing thermodynamic cycles in which thermal and quantum statistical effects coexist in a new type of hybrid quantum engine that can show significant performance boosts. 

In particular, our results show that in a box potential the effects originating from statistical nature of an ideal Bose gas and a Tonks-Girardeau gas can survive at high temperatures owing to the decreasing density of states at higher energies. This results in a peak gain in work output around the Fermi temperature. In fact, we can have shown that by changing the statistics during the cycle one can leverage the differences of bosonic and fermionic statistics in different ways to maximize performance. For the \textbf{A}-engine this ensures that the engine operates at the Cuzon-Ahlborn bound regardless of temperature, while for the \textbf{T}-engine the Carnot efficiency can be reached when the effect of quantum statistics is dominant. For completeness, we have also shown that our hybrid approach is applicable to different thermodynamic devices, such as refrigerators, indicating  promising applications beyond engines. 

Finally, we note that the engine cycles we describe could be readily applied to the Pauli engine, where the working medium uses the BEC-BCS crossover to achieve a change in statistics \cite{Koch2023}. However, here interaction effects stemming from molecule-molecule scattering and larger binding energies may affect the performance of such an engine and it would be interesting to investigate this in the future.

\acknowledgements
This work was supported by the Okinawa Institute of Science and Technology Graduate University. The authors are grateful for the the Scientific Computing and Data Analysis (SCDA) section of the Research Support Division at OIST. K.M.~is grateful to Prof.~Artur Widera and Prof.~Eric Lutz for fundamental discussions on statistics engines. 
T.F.~thanks Dr.~Momo Boubakour for stochastic exchanges. T.F.~acknowledges support from JSPS KAKENHI Grant No. JP23K03290. T.F.~and T.B.~are also supported by JST Grant No. JPMJPF2221.

\section{Appendix}
\subsection{A. Analytical Background}

The energy and the entropy of an ideal Bose gas at temperatures in the degenerate regime in a box is given by \cite{kerr2024analytic}
\begin{equation} \label{E-BG_degenerate}
    E_\text{BG} = N\frac{\hbar^2 n^2 }{2m}\frac{(\pi^2 T)^\frac{3}{2}}{2\sqrt{\pi}}\left[ \frac{1}{2} \zeta\left(\frac{3}{2}\right) 
 - \frac{\pi}{2}\frac{\frac{4}{\sqrt{\pi T}}-\zeta\left(\frac{1}{2}\right) }{\left( \frac{2}{\sqrt{\pi T}}-\zeta\left(\frac{1}{2}\right) \right)^2} \right].
 \end{equation}
 \begin{equation}
 S_\text{BG} = Nk_B \frac{\sqrt{\pi T_h}}{2} \left[ \frac{3}{2} \zeta\left(\frac{3}{2}\right)  - \frac{\pi}{2}\frac{\frac{8}{\sqrt{\pi T_h}}-3\zeta\left(\frac{1}{2}\right) }{\left( \frac{2}{\sqrt{\pi T_h}}-\zeta\left(\frac{1}{2}\right) \right)^2} \right] 
\end{equation}
while for the TG gas the energy and entropy equations for the ideal Fermi gas can be used
\begin{equation}\label{E-TG_degenerate}
    E_\text{TG} = N \frac{\hbar^2 \pi^2 n^2}{2m} \left( \frac{1}{3}+ \frac{(\pi T)^2}{12} + \frac{(\pi T)^4}{60} + \frac{35(\pi T)^6}{1296} \right)
\end{equation}

\begin{equation}
    S_\text{TG} = Nk_B \left(  \frac{\pi^2 T}{6}+\frac{\pi^2 T^3}{45} +\frac{7\pi^6 T^5}{216}  \right).
\end{equation}
In the non-degenerate regime the above equations have to be replaced by their respective high-temperature counterparts, which for the the ideal Bose gas are given by
\begin{align}\label{E-BG_nondegenerate}
   E_\text{BG} &= N\frac{\hbar^2 n^2}{2m} \Bigg[ \frac{\pi^2 T}{2} - \frac{1}{2}\sqrt{\frac{\pi^3 T}{2}}  \nonumber\\ 
  &\qquad\qquad\quad  +\pi \left( 1-\frac{4}{3\sqrt{3}}\right) + \pi^\frac{1}{2}\frac{2\sqrt{3}-5}{2\sqrt{2 T}} \Bigg].
  \end{align}
  \begin{equation}\label{S-BG_nondegenerate}
    S_\text{BG}=  Nk_B \left[ \ln \left({\frac{\sqrt{\pi T}}{2}}\right) + \frac{3}{2} + \frac{1}{2\sqrt{2\pi T}} + \pi^\frac{1}{2} \frac{2\sqrt{3}-5}{2\sqrt{2T}} \right].
\end{equation}
and for the TG gas by
\begin{align}\label{E-TG_nondegenerate}
  E_\text{TG} &= N\frac{\hbar^2 n^2}{2m} \Bigg[ \frac{\pi^2 T}{2} + \frac{1}{2}\sqrt{\frac{\pi^3T}{2}} \nonumber\\
  & \qquad\qquad\quad + \pi \left( 1-\frac{4}{3\sqrt{3}}\right) + \pi^\frac{1}{2}\frac{2\sqrt{3}-5}{2\sqrt{2T_h}} \Bigg].
  \end{align}
  \begin{align}\label{S-TG_nondegernate}
  S_\text{TG} &= Nk_B \Bigg[ \ln \left({\frac{\sqrt{\pi T}}{2}}\right) + \frac{3}{2} \nonumber\\
    &\qquad\qquad\quad- \frac{1}{2\sqrt{2\pi T}} + \pi^{1/2} \frac{2\sqrt{3}-5}{2\sqrt{2 T}} \Bigg].
\end{align}

\subsubsection{\textbf{Case 1: A-engine}}

Applying these analytical expressions to the \textbf{A}-engine the efficiency can be calculated in the degenerate regime as given by Eqs.~\eqref{analyticalBG_TG_A} and \eqref{analyticalTG_BG_A} in the main text. We note here again that the entropy needs to be conserved in the work strokes $\langle W_1 \rangle$ and $\langle W_3 \rangle$, which requires that the initial equilibrium temperature must be replaced by an effective temperature $T_h'$ for these strokes. However, as we have considered the simple case of $T_c=0$ for the compression (\textbf{a} $\to$ \textbf{b}) work stroke, the entropy is trivially conserved, whereas for the expansion (\textbf{c} $\to$ \textbf{d}) the temperature $T_h$ at point \textbf{c} will become an effective temperature $T_h'$ at point \textbf{d}, such that $S_c(T_h) =S_d(T_h') $. This allows to use the leading terms of Eqs.~\eqref{E-BG_degenerate} and~\eqref{E-TG_degenerate} for the accurate calculation of $\langle W_3 \rangle$.

The same analysis can be done for the non-degenerate regime using the high-temperature approximations for the energy, Eqs.~\eqref{E-BG_nondegenerate},\eqref{E-TG_nondegenerate} and entropy, Eqs.~\eqref{S-BG_nondegenerate},\eqref{S-TG_nondegernate}. Here, because we are only taking the leading orders of $T_h$, equating $S_c(T_h)$ and $S_d(T_h')$, returns $T_h=T_h'$. For the BG-TG cycle the high temperature equations then read as
\begin{equation}
   \eta_\text{A} \sim 1 - \left[ \frac{\left(\frac{T_h}{2} - \frac{T_h^{1/2}}{2\sqrt{2\pi} }  \right)}{\left( \frac{T_h}{2 } +\frac{T_h^{1/2}}{2\sqrt{2\pi} } - \frac{1}{3}\right)}\right]   \left(\frac{L_\textbf{b}}{L_\textbf{a}}\right)^2,
   \label{eq:A_BGTG}
\end{equation} 
and for the TG-BG cycle as
\begin{equation}
   \eta_\text{A} \sim 1 - \left[ \frac{\left(\frac{T_h}{2} + \frac{T_h^{1/2}}{2\sqrt{2\pi} }  - \frac{1}{3}\right)}{\left( \frac{T_h}{2 } - \frac{T_h^{1/2}}{2\sqrt{2\pi} }  \right)}\right]   \left(\frac{L_\textbf{b}}{L_\textbf{a}}\right)^2.
   \label{eq:A_TGBG}
\end{equation}
We have confirmed that these efficiencies, like their numerical counterparts, converge to the Otto efficiency, $\eta_\mathcal{O}$, for high temperatures of the hot bath, $T_h$.

\subsubsection{\textbf{Case 2: T-engine}}
For the BG-TG cycle in the \textbf{T}-engine the high temperature expression for the efficiency is given by
\begin{equation}
   \eta_\text{T} \sim  1 - \left[ \frac{\left(\frac{T_h}{2} + \frac{T_h^{1/2}}{2\sqrt{2\pi} }  - \frac{1}{3}\right)}{\left( \frac{T_h}{2 } - \frac{T_h^{1/2}}{2\sqrt{2\pi} }  \right)}\right]   \left(\frac{L_\textbf{b}}{L_\textbf{a}}\right)^2,
   \label{eq:T_BGTG}
\end{equation} 
and for the TG-BG cycle
\begin{equation}
   \eta_\text{T} \sim  1 - \left[ \frac{\left(\frac{T_h}{2} - \frac{T_h^{1/2}}{2\sqrt{2\pi} }  \right)}{\left( \frac{T_h}{2 } +\frac{T_h^{1/2}}{2\sqrt{2\pi} } - \frac{1}{3}\right)}\right]   \left(\frac{L_\textbf{b}}{L_\textbf{a}}\right)^2.
   \label{eq:T_TGBG}
\end{equation}
Again, these converge to the Otto efficiency, $\eta_\mathcal{O}$, for high temperatures of the hot bath, $T_h$. We also note that Eq.~\eqref{eq:A_BGTG} is the same as Eq.~\eqref{eq:T_TGBG}, and Eq.~\eqref{eq:A_TGBG} is the same as Eq.~\eqref{eq:T_BGTG}.

\subsection{B. \textbf{GV} cycle efficiency}

For the calculation of the efficiency of the \textbf{GV} cycle, we need to only compute the work $\langle W_1\rangle$ and $\langle W_3\rangle$ and the total heat input, which for this reduced \textbf{T}-engine would be $\langle Q_2\rangle_\text{Total} = \mathcal{E}^{(\textbf{d})} -  \mathcal{E}^{(\textbf{b})} $. This gives the expression for the efficiency for the BG-TG cycle as
\begin{equation}
    \eta_\textbf{GV} = -\frac{\sum_{ n} b^{(\textbf{a})}_{n}\left[E^{(\textbf{b})}_{n}-E^{(\textbf{a})}_{n}\right] + \sum_{n} f^{(\textbf{d})}_{n}\left[E^{(\textbf{a})}_{n}-E^{(\textbf{b})}_{n}\right]}{\sum_{n} \left[f^{(\textbf{d})}_{n} - b^{(\textbf{a})}_{n} \right]E^{(\textbf{b})}_{n}}.
\end{equation}  
This equation can be further simplified to
\begin{equation}
    \eta_\textbf{GV}  = 1 - \frac{\sum_{n} \left[f^{(\textbf{d})}_{n} - b^{(\textbf{a})}_{n} \right] E^{(\textbf{a})}_{n}}{\sum_{n} \left[f^{(\textbf{d})}_{n} - b^{(\textbf{d})}_{n} \right]E^{(\textbf{b})}_{n}}
\end{equation} 
and due to scale invariance, $E_{n}^{(\textbf{a})}/E_{n}^{(\textbf{b})} = (L_\textbf{b}/L_\textbf{a})^2$, the above efficiency reduces to the Otto efficiency. In the same way the above calculation for the TG-BG cycle also gives an efficiency equal to the Otto efficiency.

 \bibliography{Reference}

\end{document}